# A Vision of Nuclear and Particle Physics

Hugh E. Montgomery
*Jefferson Lab, Newport News, VA 23606, USA*

Abstract

This paper will consist of a selected, personal view of some of the issues associated with the intersections of nuclear and particle physics. As well as touching on the recent developments we will attempt to look at how those aspects of the subject might evolve over the next few years.

## 1. INTRODUCTION

The brief for this paper was to provide a vision of nuclear and particle physics. Perforce, given limited time, this meant making choices; even after making the choices, brutal triage was applied and the result is a very select number of topics and even those will be treated very superficially. I am clearly guilty of bias; many of the choices favor my interests either personal or institutional.

There have been several planning exercises during the past few years in both particle [1,2], and nuclear [3,4] physics. While it is permitted to dream about experiments that could happen, in the main, we would expect that things which do happen will have been considered and supported by these studies.

I have chosen to partition the paper as follows. In Section 2, I will discuss two manifestly energy frontier subjects, the top quark and the Higgs. In Section 3, I will address Hadron Physics, a subject in which is on the boundary between nuclear and particle physics, and in Section 4, I would like to touch on some of the impressive advances made in the past decade in Lattice Gauge calculations of Quantum Chromodynamics (QCD). The final subject, in Section 5, will look to measurements related to the fundamental symmetries of nature, before we try to make a few conclusions in Section 6.

## 2. THE HIGHEST ENERGIES

The highest energies for some time to come will be provided by the Large Hadron Collider (LHC) at CERN. Here the number of topics could have easily occupied several talks. I have chosen two of the topics. I have known the top quark since before it officially existed, and the Higgs particle sits at the apex of our current understanding of particle physics.

Progress in nuclear and particle physics is strongly influenced by progress in technology, and, in particular, with the accelerators. This was certainly true in the discovery of the top quark at the Tevatron; no other machine in the world had the energy and the experiments were ready, just waiting for the luminosity. It was true again in the discovery of the Higgs. In that case the Tevatron had come close but not close enough, while the LHC delivered what was needed. The prospects going forward are clear. We expect about a factor of ten in luminosity and in integrated luminosity going from Run 1 to Run 2 of the LHC. What will be most important, especially for the discovery physics, is the factor of two increase in energy. Going from 1800 GeV to 1960 GeV at the Tevatron the high mass production cross sections went up by 40%; with the factor of two, from 7 TeV to 13(14) TeV in energy at the LHC, the increases will be much more dramatic.

## 2.1. The Top Quark

The top quark was twenty years old this spring. It is amusing to look at how the data available on the mass of the quark has evolved. Mass plots in '95 were quite rudimentary, but very quickly it became most accurately measured of any of the quark masses. I will just concentrate on the mass, although there are other important properties which will see substantial improvement. At present the mass data are already very precise, and the Tevatron precision is still competitive with an uncertainty of about 1 GeV (0.35%).

The expectations [5], see Figure 1, are that during Run 2 CMS and ATLAS uncertainties will dip substantially below that level, likely getting to 0.5 GeV. The situation will reach a point at which the issue becomes not the experimental precision per se but rather the relationship between the measurement and a theoretically well-defined mass measure (the pole mass for example). Are the QCD corrections to the mass which is measured under control or not?

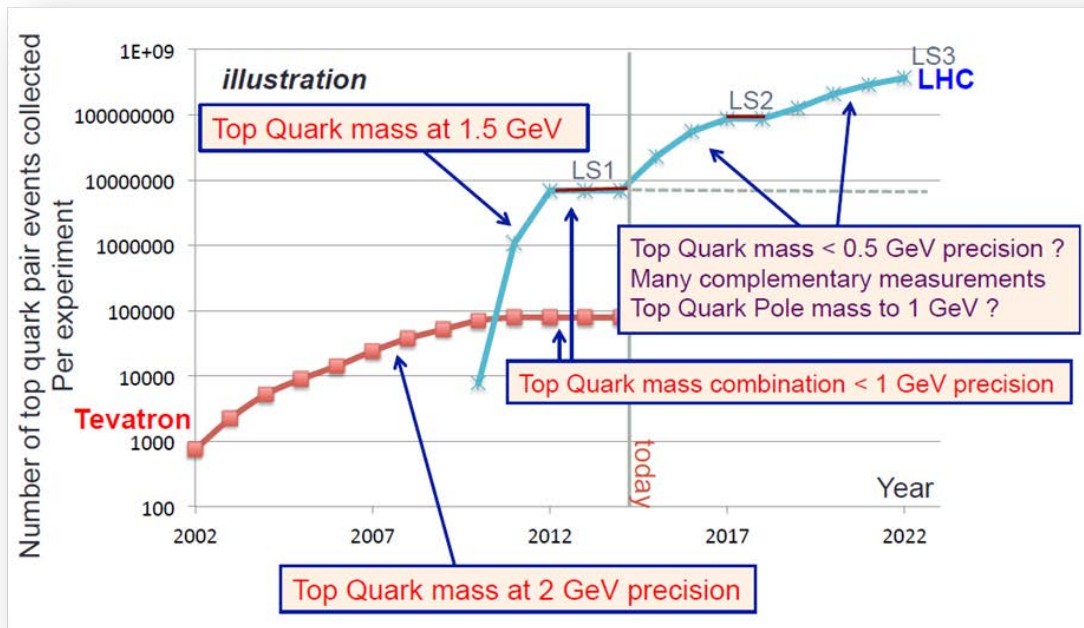

Figure 1: Recent history of the top quark and milestones in its mass measurement with projections through the next decade of Large Hadron Collider running.

## 2.2. The Higgs Boson

At the time of observation [6,7] of a phenomenon in two independent experiments, the first question is whether they each have observed the same object. This then leads to the question "What have they observed?" Since we have sought the Higgs for so long our duty is to be skeptical. The yields of Higgs in the different channels so far observed appear to follow expectations as a function of mass, however, they still have significant statistical uncertainties. Currently, these range from 20% for the gamma-gamma and the W and Z channels to something in excess of 50% for the b-quark channel. In Run 2 however, we expect that even the difficult b-meson and τ-meson channels will be measured with an uncertainty less than 15% [8,9].

With the evidence in hand, ATLAS and CMS have observed the Higgs Boson, with spin-parity $0^+$, and mass 125 GeV. The goal for Run 2, one way or another will be to find new physics. This could come in the form of some kind of partner for the Higgs boson, or it could come through the discovery of some other kind of particle, a harbinger of super-symmetry or technicolor, or it could come in the behavior of boson-boson scattering at the very highest center of mass energies.

If new physics does not manifest itself in Run 2, then the importance of alternative routes to discovery, the virtual path through measurements of rare processes and high precision measurements of fundamental parameters with lower energy machines, will be enormously enhanced.

## 3. HADRON PHYSICS

A major player in Medium Energy physics is Jefferson Lab with its Continuous Electron Beam Accelerator Facility (CEBAF). The 12 GeV upgrade project more than doubles the kinematic phase space which can give access to deep inelastic scattering and thus the region in which the partonic approach, to the theoretical calculations, is valid. However, note also the considerable flexibility in luminosity and polarization; the use of polarized targets is also ubiquitous. The project is more than 95% complete, the accelerator is operational, the Hall D construction, including the GlueX experiment is complete, and its commissioning is well advanced. There is also good progress with the challenging superconducting magnets needed for the new spectrometers in Halls B (CLAS 12) and C (SHMS). Other important medium energy hadron physics programs are the Compass muon and hadron scattering experiment at CERN, Seaquest, the Drell-Yan program at Fermilab, and the RHIC-Spin program at Brookhaven.

### 3.1. Spectroscopy

For some time, the participation of gluons in the development of the quantum numbers of hadronic states has been anticipated. Hadronic states with glueball characteristics or hybrids with the valence structure involving gluons have long been sought. The GlueX experiment, at Jefferson Lab, will launch a new campaign. It is hoped that the use of a polarized photon beam will permit the identification of some hybrid states whose quantum numbers cannot be constructed using only quarks; gluons are necessary. Lattice gauge calculations relevant to this search will be discussed later. Note that the CLAS 12 facility may also get to play in this game.

### 3.2. Structure of the Nucleon

The distributions of partons as a function of momentum fraction, Bjorken-x, and the studies of form factors of nucleons are essentially one-dimensional representations of the internal structures of hadrons. These distributions result from integrating over all but one of the five dimensions of the quantum mechanical phase space, Wigner, distribution, Figure 2. Based on the work of Xiangdong Ji [10], and Anatoly Radyushkin [11], more differential descriptions were developed. The descriptions, involving both longitudinal and transverse distributions, introduce Generalized Parton Distributions (GPDs) and Transverse Momentum Distributions (TMD). Exploratory studies have come from the HEMES experiment, at DESY, from COMPASS, and from Jefferson Lab. At Jefferson Lab some of the earliest experiments with 12 GeV are opening the new campaign to explore the three-dimensional structure of nucleons and nuclei.

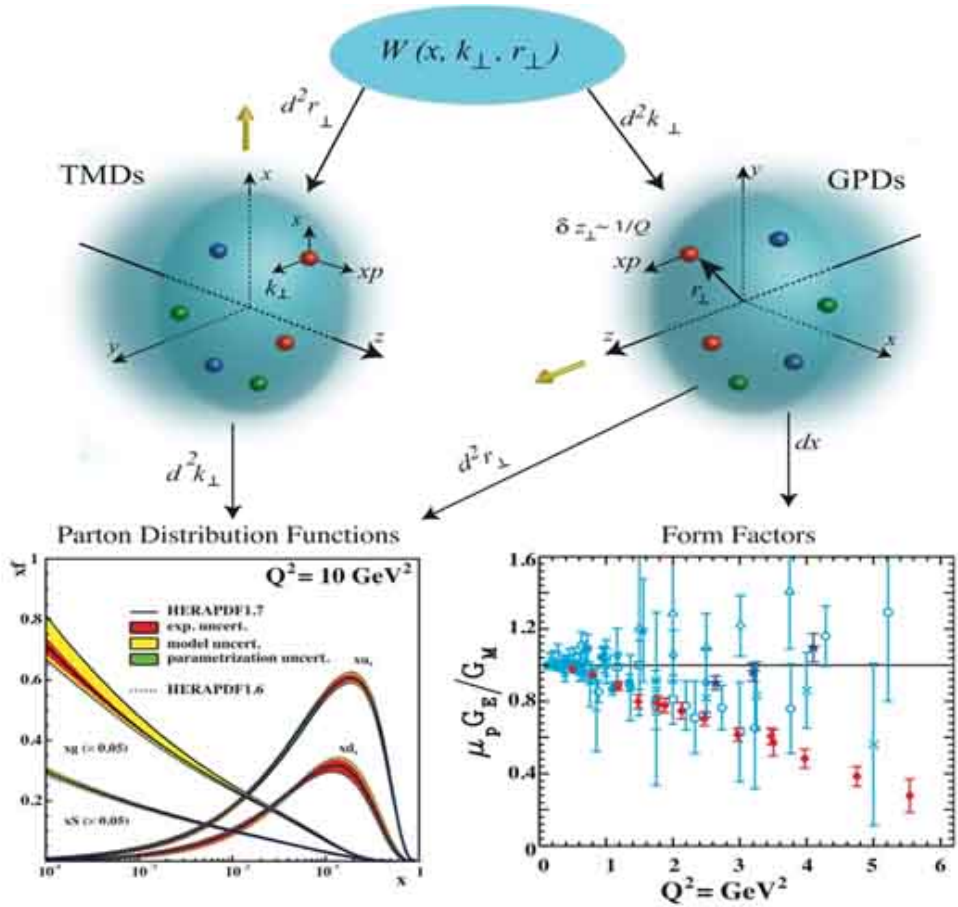

Figure 2: Relationships of parton distributions and form factors to the Wigner distribution of partons.

### 3.3. Connections to the Nucleus

While the EMC effect first came to light in 1982 [12], a complete understanding of the phenomenon is still sought; it represents a significant fraction of the Jefferson Lab program [13]. Contemporary data relate the slopes of the EMC effect curves to the local nuclear density and to the emergence of multiple nucleons in the scattering process. The latter phenomenon makes an inescapable link to the short range nuclear force. The gross imbalances in the numbers of (pn) as compared to (pp) and (nn) pairs, which result, suggest tensor effects. What is missing is a good coherent description of these multiply linked phenomena. I am sure we will see more experimental data, but I am not yet convinced that we will see enlightenment.

The features of the Jefferson Lab accelerator, its high beam current, and its control of beam helicity permit the use of parity violating electron scattering (PVES) as a basic tool. One example is PVES on a lead target. The Z boson, with which the photon exchange interferes, has a stronger, by far, coupling to the neutron than to the proton. So PVES "sees" the neutron distribution. There is a skin of neutrons beyond the protons. This has been observed by the PREX experiment [14]; the measurement will be repeated with higher precision. An upside risk to this technique is that it could turn into a cottage industry; there are a lot of nuclei and already a measurement with $^{48}$Ca has been approved.

I have concentrated on physics that will be done with CEBAF in the next decade with 12 GeV beams. Of course the medium energy community has ambitions for more. QCD is a very rich field so we can imagine further advances in thinking analogous to the GPD and TMD constructs. But, in particular, the extension of all these beautiful studies from the valence region to the low x gluon dominated region is attractive. Both the BNL and Jefferson Lab communities have been looking hard at this and have generated a common physics case [15]. The two laboratories have their individual collides designs [16,17], which exploit their existing infrastructures. There is hope that that the recently completed, soon-to-be released, NSAC long range plan, will recommend an Electron Ion Collider for construction in the 2020s.

## 4. THE LATTICE

Computing is very important of course and we see the influence in the reductions of the effective pion mass in the calculations and the improvements in precision of determination of some of the classic parameters. However, also important has been the expansion of the operator space used, which has enabled qualitatively different lines of investigation. So, recent advances in lattice gauge calculations have not been predicated solely on the advances in computing power.

### 4.1. Heavy Quarks

The charmonium and bottomonium states well below threshold for meson production involve a single channel and were well treated a number of years ago. However the introduction of more interpolating operators corresponding to multiple final states has led to descriptions [18] of states such as the ψ(3770) which is just above threshold for D-Dbar decay. The technique of introducing these extra interpolating operators also produced [19,20] a lattice candidate state in charmonium, which is a good candidate explanation for the X(3872) state observed in several experiments. We anticipate further extensions of these techniques to quarkonium physics.

### 4.2. Light Quarks

The necessity to consider multiple final states and nearby thresholds has meant that successful descriptions of the light quark meson spectra have only recently emerged, but the advances are now very exciting; Figure 3 shows the isospin-1 spectrum [21,22]. Features include a rather complete spectrum of excited states, including a super-multiplet of hybrid states, including some with exotic quantum numbers. Their masses are about 1.2 GeV above the ρ-meson. The suggestion that the gluons are playing a valence-like role is clear.

### 4.3. Interactions

A final ingredient needed to describe hadron decay spectra is an understanding of inelastic interactions. A couple of years ago, lattice calculations could identify the phase shift associated with the ρ-meson in the elastic π−π channel [23]. And now we see the presence of both πK and the inelastic ηK channels in πK scattering [24]. The technology at work here, the broad set of interpolating operators as for the heavy quark states above threshold, is related to that used to describe the light quark spectrum.

A decade from now, I would expect that the discussions about how close to the pion mass we will have reduced to almost zero, and the issue will be the complexity which can be handled.

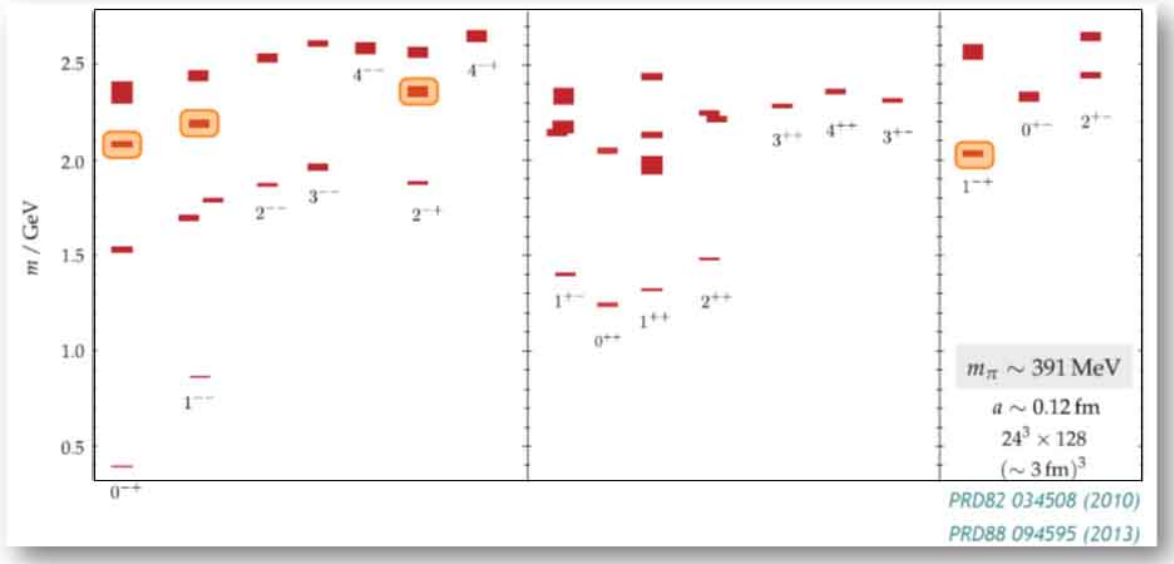

Figure 3: Isospin-1 meson spectrum from lattice gauge calculations. A possible super multiplet of hybrid mesons [$(0,1,2)^{-+}$, $1^{--}$], with masses approximately 1.2 GeV above that of the ρ meson is indicated by the shaded boxes.

## 5. SYMMETRIES

The study of symmetries is fundamental to many branches of physics, not least to both particle and nuclear physics. In the spirit of giving credit to machines and techniques, let me again call attention to PVES. This approach made a spectacular entry into the world of physics with the Prescott measurement [25] in the 1970s of the scattering from deuterium. This measurement of $\sin^2\theta_W$ was very importantly a beautiful demonstration of the unification of electromagnetic and weak interactions. Today, the technique has matured considerably which has led to multiple uses. For example, it was used to demonstrate that the strange quark content of the nucleon is very small. Earlier we discussed the measurement of the neutron distribution in nuclei. Here we concentrate on the mixing angle determination. A recent measurement [26] of the weak charge, Qweak, of the proton is still in analysis, the results for the full data set are eagerly awaited; it is the first of a series of similar measurements.

### 5.1. Electroweak

A Moller scattering measurement [27] with an experiment of the same name is being developed at Jefferson Lab. It is the successor to E158 [28] at SLAC and has a beautifully novel design. The toroidal magnets have odd numbers of coils (7) so that for ee scattering, with two identical particles in the final state, the acceptance can be 100% in azimuth, despite the coils. The dimensions, 30 m. long by 1-2 m. in radius give the experiment a novel aspect ratio. The current value of $\sin^2\theta_W$ would give a measurable asymmetry of 35 parts per billion; the goal for the experiment is to achieve a precision of less than 1 part per billion, a measurement uncertainty of 2% (stat) and 1% systematic in $\sin^2\theta_W$. Note that the precision will be comparable to that of the best LEP or SLC measurements. But note also the recent measurement [29] from Dzero, which is also comparable, and the anticipated legacy value from the Tevatron. There are other ambitions for a low energy measurement at Mainz. As mentioned earlier, the importance of low energy precision measurements may well be increased in the continued absence of new-physics clues from LHC.

### 5.2. Neutrinos

Exploration of neutrino physics has often uncovered apparent anomalies, some of which led to discoveries. The neutrinos were found to have mass. Immediately the questions multiplied and have resulted in an enormously productive couple of decades and presage a couple more to come.

Neutrino physics uses multiple sources, the sun, cosmic rays, accelerators and reactors. It is important to know the characteristics of the source and this has not been easy. The "near detector" is an effort in many experiments to take some of the dead reckoning out of the enterprise. However, a series of short baseline experiments has also thrown up its own sequence of not-quite-resolved-yet anomalies which allow for speculation about species of neutrinos beyond our, now familiar, three. A very recent addition to the line-up is Micro-BooNE [30], which also has the mission of establishing Liquid Argon technology in the "new world". Looking to the future there is a lot of activity anticipated at Fermilab with the advent of an experiment based on a transformed ICARUS [31] device of Gran Sasso fame. One way or another the array of near detectors on the Fermilab site is expanding. The ambitions should be to seek clarity with respect to sterile neutrinos but importantly, to enable the long baseline measurements.

Recent long baseline experiments have made significant contributions to neutrino physics with K2K based on the KEK accelerator, MINOS which used the NuMI beamline at Fermilab, and with the T2K beam from the JPARC accelerator. A new experiment [32], which started operating in the past year, is the NOvA experiment "off-axis" from the NuMI beamline. The T2K experiment was the first to observe the $\nu_\mu$ to $\nu_e$ transition. The Daya Bay reactor experiment has since set the gold standard for the determination of the relevant mixing parameter, $\theta_{13}$.

The figure of merit for the accelerators is power. T2K aims to provide 750 kW eventually; the Main Injector at Fermilab is producing about 4-500 kW and expects to reach 700kW in 2016. The expected sensitivities for the mass hierarchy and for the determination of the CP violating phase angle are shown in Figure 4 for the combined NOvA and T2K experiments.

In the farther future, the Deep Underground Neutrino Experiment, DUNE [33], based on a largely improved neutrino source at Fermilab is being planned as a preeminent international initiative in the field. There also remains the possibility that, if the HyperKamiokande initiative [34] is realized, they will have competition.

Kamland, and more recently, the Daya Bay experiment, demonstrated, in unequivocal terms, how a well-designed reactor experiment can dominate the field. The goal of the JUNO experiment [35] is to look for the disappearance of electron-type antineutrinos which come from the nuclear reactors. There are major challenges, in particular the achievement of the energy resolution, but, if successful, the experiment could provide a clear determination of the mass hierarchy.

Finally let me mention the growth of the program to seek neutrinoless double beta decays of various elements. The next generation experiments will have sensitivity [3] to half-lives of order $10^{26}$ years. These experiments have the goal of determining whether or not the neutrino mass derives from the Dirac or Majorana nature of the particles. The ambitions are to make measurements which are sensitive to Majorana masses of 10 millielectron volts. The build up to a technology choice will be challenging, but we expect such experiments to be operational about a decade from now.

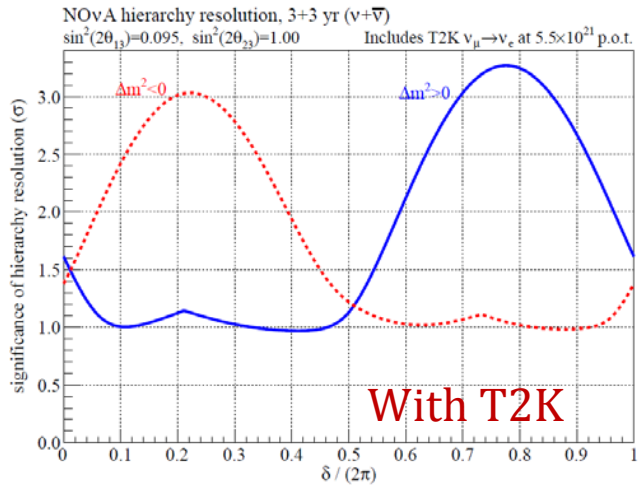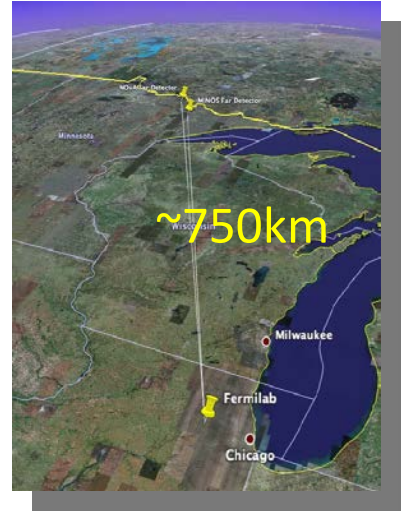

Figure 4: Left: anticipated sensitivity of the NOνA experiment in conjunction with the T2K experiment. Right: footprint of the NOνA experiment.

### 5.3. Dark Matter

The experimental basis for dark matter rests on a number of astronomical measurements which are quite compelling. The only thing missing is the direct observation; there were some years between the time when the neutrino was postulated and its direct detection. However, it is now the better part of a century since the first seeds of dark matter were sown and the searches have not borne fruit. The advances in technology and mass, though, have resulted in limits many orders of magnitude more sensitive in the space of a decade and also considerable extension of mass range. Detectors are under construction [36] which could be sensitive to coherent neutrino scattering, which is a possible floor for the experiments. Certainly the measurements are stretching the comfortable range for models such as Supersymmetry.

### 6. CONCLUSION

An inescapable comment is that the standard model continues to provide a remarkable description of the fundamental aspects of particle physics; this is exemplified by Figure 5 [37] which demonstrates the mutual consistency of many electroweak measurements. In the hadron arena, we are making good progress in understanding the enormously rich physics which derives from the very simple tenets of quantum chromodynamics. On the discovery side, our knowledge of neutrinos has seen spectacular progress, but not all aspects have yet been explored, so this must represent a primary opportunity. There are other places where our instinct has led us to invest major effort, and they still seem attractive. The dark matter search program seems well founded; but, we are puzzled by our continued lack of revelation. The LHC will continue to be our spearhead, but I have inserted a couple of comments in this paper that the high precision, lower energy experiments may assume a greater importance than ever, depending on what we find in Run 2. In all the fascination of nuclear and particle physics will remain.

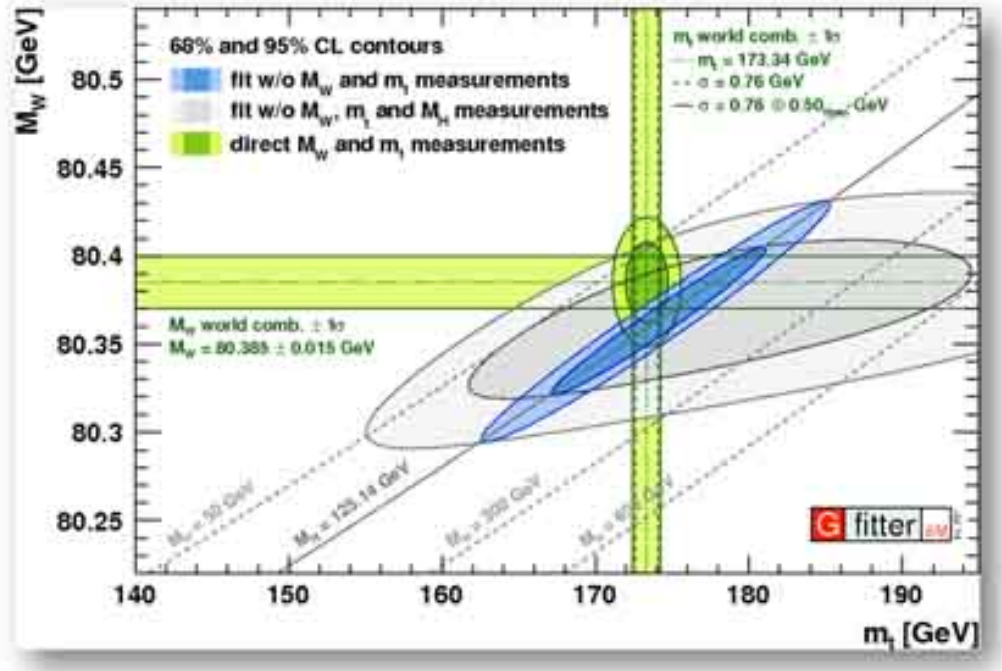

Figure 5: The classic figure showing the various electroweak measurements on a plot of the mass of the top quark versus the mass of the W boson.

## Acknowledgments

I appreciate input and conversations over the past couple of years from man colleagues: Robert Edwards, Rolf Ent, Krishna Kumar, Andy Lankford, Nigel Lockyer, Bob McKeown, Meenakshi Narain, Mike Pennington, Sasa Prelovscc, and Steve Ritz. Meenakshi pointed me at presentations including several from a number of younger physicists: The Highest Energies: Aram Apyan, ECFA 2014; Si Jie, Moriond 2015; Hubert Kroha, Aspen 2015; Meenakshi Narain; Top at Twenty, 2015; Andrew Whitbeck, LPC 2014; and the public ATLAS and CMS sources. I leaned on the presentations from my Jefferson Lab colleagues Hadron Physics: Rolf Ent, Kitty Hawk 2015, and Lattice: Robert Edwards, NSERC 2014. For Symmetry: Krishna Kumar, Kitty Hawk 2015; Nigel Lockyer, HEPAP 2015; Bob McKeown, NSAC 2014; PANIC 2014; Y.F. Wang; and Neutrino Web Pages provided the material. This work was supported by DOE contract DE-AC05-06OR23177, under which Jefferson Science Associates, LLC, operates the Thomas Jefferson National Accelerator Facility.